# NMR method for amplification of single spin state

Gregory B. Furman[a,b], Victor M. Meerovich[a], and Vladimir L. Sokolovsky[a]

[a] Physics Department, Ben Gurion University, P.O.B. 653, Beer Sheva, 84105, Israel

[b] Ohalo College, P.O.B. 222, Qazrin, 12900, Israel

Email: gregoryf@bgu.ac.il

Amplification of a single spin state using nuclear magnetic resonance (NMR) techniques in a rotating frame is considered. The main aim is to investigate the efficient of various schemes for quantum detection. Results of numerical simulation of the time dependence of individual and total nuclear polarizations for $1D$, $2D$, and $3D$ configurations of the spin systems are presented.

# 1. Introduction

Concerning the impellent problem in quantum information processing is amplifying and then measuring the output of a protocol. Recently many schemes have been proposed for realizing of measurements of a single quantum state [1-7]. These schemes will play an important role in building up a quantum device for amplification of a very low signal from a single quantum object. To estimate the efficiency of these schemes several values can be used. One of them is the contrast, $C$, introduced in Ref. [3, 4]:

$$C = \frac{M_z^{(0)} - M_z^{(1)}}{M_z(0)}, \qquad (1)$$

where $M_z^{(0)}$ and $M_z^{(1)}$ are the nuclear magnetization of the system obtained when the measured nuclear spin is in the state $|0\rangle$ or $|1\rangle$, respectively, $M_z(0)$ is the initial nuclear magnetization. Quantity for contrast received within the framework of the realizable models lies in a range of $1.6 \div 1.7$, while the maximum theoretical contrast is $2$ [1, 3, 4].

Recently, it was demonstrated a principle for quantum detection of state of a single spin state in a one-dimensional Ising chain with nearest-neighbor interactions [1] and in more realistic spin system include interactions beyond the nearest neighbors and natural dipole-dipole interactions [8]. In this model, a wave of flipped spins can be triggered by a single spin flip results that a polarization of the single spin is converted into a total polarization of the spin system. This process can be described by using a sequence of quantum gates [1, 5]:

$$CNOT_{N,N-1}...CNOT_{3,2}CNOT_{2,1}, \qquad (2)$$

where $CNOT_{n,m}$ is the control-not gate, which leads to flip the state of the $m$-th qubit when the $n$-th qubit is in the state $|1\rangle$ and does not do anything when the $n$-th qubit is in the state $|0\rangle$. The equivalent explanation can be defined by using the following physical arguments. The effective field on each spin is determined as a vector sum of the local field, $\omega_d$, produced by neighbor spins, and radio-frequency field (RF). When a spin, oriented along Z axis and irradiated by a weak RF field along X axis, $\vec{H}_1 \| X$, has the two neighbors in different states, for example, one up, along an external magnetic field, $\vec{H}_0 \| Z$, and another down, anti-parallel to $\vec{H}_0$, the local field produced by these nearest-neighbors on the spin equals to zero. The spin starts to rotate around the direction of the weak RF field and changes its initial direction. In contrast, when a spin has two neighbors in the same state, for example along $\vec{H}_0$, the local field on this spin is not zero in the $Z$ direction. The effective field coincides with the local one result that the spin remains in the former position. Evidently, the above mentioned explanation is correct for the simple model of spin system with a weak dipolar coupling between nearest neighbors when a Hamiltonian includes only the $Z$ components of the spin operators. In real solids, the main interaction is a dipole-dipole one with coupling all spins in a cluster, and the Hamiltonian includes all components of spin operators. The previous proposed methods for single spin measurements concentrate on the linear chain (1D) [1, 8] and cubic (3D) [6] spin systems.

In the present paper we compare the efficiency of the 1D, 2D, and 3D configuration of the spin systems to realize a single spin measurement scheme. The advantage of 2D and 3D configurations over the previous schemes [1, 6, and 8] is that all spins have the same possibility to be flipped simultaneously. We consider a nuclear spin system in a rotating frame representation. The system consists of two parts. One of the parts is simple one spin (*S*), state of which is detected. The second part is a system of dipolar coupling homonuclear spins, which plays role in the measuring device. Our main goal is convert the polarization of the target single spin (*S*) into a total polarization of the spin system. First we consider one-dimensional spin systems, namely, chains and rings of spins in the limit of weak coupling [9]. Then the amplification processes are studied in two- and three- dimensional spin systems. Finally, we discuss the efficiency of these schemes. We suggest that internal spin dynamics, with natural dipole-dipole interaction, can serve as a "pre-detector" for producing stronger signals and may even help in reaching the ultimate limit of a single-spin detection by using clusters of coupled nuclear spins and conventional NMR techniques and spectrometers.

## 2. Average Hamiltonian

Let us consider the model of a system of nuclear spin $S$ and $N$ spins $I$ coupled by dipole-dipole interaction, placed in a high magnetic field $\vec{H}_0$ directed along the $Z$ axis and irradiated on resonance by weak transverse RF fields, $\vec{H}_1(t) = \vec{H}_1 \cos\Omega_0 t$ and $\vec{h}_1(t) = \vec{h}_1 \cos\omega_0 t$, where $H_1$ and $h_1$ are the amplitudes of the weak RF fields. The Hamiltonian of the spin system can be presented in the following form:

$$H_{lab} = \Omega_0 S^z + \Omega_1 S^x \cos\Omega_0 t + \omega_0 \sum_{n=1}^{N} I_n^z + \omega_1 \sum_{n=1}^{N} I_n^x \cos\omega_0 t + H_{dd} + H_{SI} \qquad (3)$$

where $\Omega_{0,1} = \gamma_S H_{0,1}$, $\omega_0 = \gamma_I H_0$ and $\omega_1 = \gamma_I h_1$, $\gamma_S$ and $\gamma_I$ are the gyromagnetic ratio of nuclei $S$ and $I$. $S^z$, $S^x$, $I_n^z$ and $I_n^x$ are the angular momentum operators of $S$ and $I$ nuclei in the $Z$ and $X$ directions, respectively. $H_{dd}$ is the secular part of the dipole-dipole interaction Hamiltonian with the coupling constant $d_{mn}$ between $I$ nuclei

$$H_{dd} = \sum_{m>n} d_{mn} \left[ I_m^z I_n^z - \frac{1}{2}\left(I_m^x I_n^x + I_m^y I_n^y\right) \right]. \tag{4}$$

and $H_{SI}$ is the secular part of the dipole-dipole interaction Hamiltonian with the coupling constant $g_m$ between unlike spins $I$ and $S$ nuclei

$$H_{SI} = \sum_m g_m S^z I_m^z. \tag{5}$$

In the rotating frame at [10, 11] $\Omega_0 \approx \omega_0 \gg d_{mn} \approx g_m \gg \omega_1 = \Omega_1$, the fast oscillating terms can be removed [10, 11]:

$$H_{rot} = \frac{\Omega_1}{2} S^x + \frac{\omega_1}{2} \sum_{n=1}^{N} I_n^x + H_{dd} + H_{SI}. \tag{6}$$

To reach our goal and convert the polarization of the spin $S$ to the total polarization of the spin system $I$ the Hartmann-Hahn condition [12], $\omega_1 = \Omega_1$ must be fulfilled.

Usually, at $\omega_1 = \Omega_1 \gg \omega_d = \sqrt{\dfrac{Tr(H_{dd})^2}{Tr\left(\sum_n I_n^z\right)^2}}$, the dipolar Hamiltonian, $H_{dd}$, can be taken into account by averaging procedure [10, 11]. The case is considered here is opposite, $\omega_1 = \Omega_1 \ll \omega_d$. To correctly take into account the first two terms in the right-hand side of Eq. (6) let us perform the unitary transformation of the first two terms of the Hamiltonian (4) [1]:

$$\widetilde{H}(t) = \frac{\omega_1}{2} U(t)\left( S^x + \sum_{n=1}^{N} I_n^x \right) U^+(t), \quad (7)$$

where

$$U(t) = \exp[-it(H_{dd} + H_{SI})]. \quad (8)$$

At $\omega_d \gg \omega_1$ in the lowest order the time-independent part of the Hamiltonian (7), the effective time-independent Hamiltonian can be obtained [10, 11]

$$H_{eff}^{(0)} = H^{(0)} = \frac{1}{t_c} \int_0^{t_c} dt \widetilde{H}(t). \quad (10)$$

### 3. Quantum state detection

Let us consider the ensemble of $N+1$ spins consists of $N$ spins $I = 1/2$ and a single spin, $S = 1/2$. The $I$-spin system will be considered as a measuring device while the $S$ spin will be regard as a target single spin. First, the $I$-spin system is prepared in a highly polarized state with the initial density matrix:

$$\rho(0) = \otimes_{k=1}^{N} \rho_k(0), \quad (11)$$

where $\rho_k(0) = \begin{pmatrix} 1 & 0 \\ 0 & 0 \end{pmatrix}_k$ is the initial density matrix of the $k$-th spin oriented *up*, along the external magnetic field $\vec{H}_0$. The method of creating the highly polarized spin states in 1D, 2D, 3D clusters of coupled spins was described previously [13, 14]. It is based on filtering multiple-quantum coherence of the highest order, followed by a time-reversal period and partial saturation. Below we consider "measuring" spin systems design as one-, two-, and three- dimension systems (Fig. 1).

### a. One-dimensional model

Let us consider a one-dimensional linear spin chain (see Fig.1a), which can be represented by spins of quasi-one-dimensional hydroxyl proton chains in calcium hydroxyapatite $Ca_5(OH)(PO4)_3$ [15]. In the limit of the weak coupling [8] the Hamiltonian describes the spin system can be presented by Eq. (3) with

$$H_{dd} = \sum_{n<m}^{N} d_{mn} I_m^z I_n^z, \qquad (12)$$

where $d_{mn} = d_1/(m-n)^3$, here $d_1$ is the coupling strength between the nearest $I$ spins. In the rotating frame Eq. (3) becomes [10, 11]

$$H_{rot} = \frac{\omega_1}{2}\left(S^x + \sum_{n=1}^{N} I_n^x\right) + \sum_{n<m}^{N} d_{mn} I_m^z I_n^z + \sum_{m}^{N} g_m S^z I_m^z, \qquad (13)$$

where $g_m = g_1/m^3$, $g_1$ is the coupling strength between the nearest $S$ and $I$ spins. The Hamiltonian (13) can be rewritten as

$$H = \frac{\omega_1}{2}\left(S^x + \sum_{n=1}^{N} I_n^x\right) + \sum_{q=1}^{M}\left(d_q \sum_{k=1}^{N-q} I_k^z I_{k+q}^z + g_q S^z I_q^z\right), \qquad (14)$$

where $d_q = d_{k,k+q}$ are the coupling constants between $k$-th and $(k+q)$-th nuclear spins and $M$ is a number of directed coupling spins. For example, at nearest neighbor approximation $M=1$, and for next-nearest neighbors approximation $M=2$. At $\omega_1 \ll d_q$ the first term in the right side of Eq. (14) can be taken into account correctly by using averaging procedure [1]. It is useful for this purpose to transform the first term of Hamiltonian (14) by the unitary transformation:

$$\tilde{H}(t) = \frac{\omega_1}{2} e^{\left[-it\sum_{q=1}^{M}\left(d_q\sum_{k=1}^{N-q}I_k^z I_{k+q}^z + g_q S^z I_q^z\right)\right]} \left(S^x + \sum_{n=1}^{N} I_n^x\right) e^{\left[it\sum_{q=1}^{M}\left(d_q\sum_{k=1}^{N-q}I_k^z I_{k+q}^z + g_q S^z I_q^z\right)\right]}. \tag{15}$$

When the modulation is very rapid compared to the magnitude, $d_q \gg \omega_1$, it becomes appropriate to take into account only the time-independent part of the Hamiltonian (15) [10, 11]. Using the transformation rules for the $X$ component of the spin operators

$$e^{-itaAB} Q e^{itaAB} = Q\cos\left(\frac{at}{2}\right) + 2PB\sin\left(\frac{at}{2}\right) \tag{16}$$

and for the $Y$ component of the spin operators

$$e^{-itaAB} P e^{itaAB} = P\cos\left(\frac{at}{2}\right) - 2QB\sin\left(\frac{at}{2}\right), \tag{17}$$

where $a = d_q$, $A = I_k^z$, $S^z$; $B = I_{k+q}^z$, $I_q^z$; $Q = I_k^x$, $S^x$; $P = I_k^y$, $S^y$, the time independent Hamiltonian in the approximation of the nearest-neighbor interaction spins $(M = 1)$ can be obtained

$$\begin{aligned}
H_{eff}^{(M=1)} &= \left(\frac{g}{2}\right) S^z + \frac{\omega_1}{8} S^x \left(1 - 2I_1^z\right) \\
&- \left(\frac{d_1 + g_1}{2}\right) I_1^z + \frac{\omega_1}{8} I_1^x \left(1 - 2I_2^z\right)\left(1 + 2S^z\right) \\
&+ \left(\frac{d_1}{2}\right) I_N^z + \frac{\omega_1}{8} I_N^x \left(1 - 2I_{N-1}^z\right) + \frac{\omega_1}{4} \sum_{k=2}^{N-2} I_k^x \left(1 - 4I_{k-1}^z I_{k+1}^z\right)
\end{aligned} \tag{18}$$

The effective time-independent Hamiltonian (18) has the following interpretation: the $k$-th spin is still at his initial state when its two neighbors are in the same state. In this case the ensemble average $\langle 1 - 4I_{k+1}^z I_{k-1}^z \rangle = 0$. This means that the local field produced by the $k-1$-th and by the $k+1$-th is doubled. When the two neighbors of the $k$-th spin are in the different states, the local field equals to zero. As results, the $k$-th spin starts to rotate around the direction of the weak RF field and changes its initial direction. The result (18) is similar to that obtained early [1], except the terms with spin operators of the first and the last spins of the chain. In the case with nearest and next-nearest neighbor interactions $(M=2)$ we obtained the following expression for the effective Hamiltonian

$$H_{eff}^{(M=2)} = \left(\frac{g_1-g_2}{2}\right)S^z + \left(\frac{d_2-d_1-g_1}{2}\right)I_1^z + \left(\frac{d_2-g_2}{2}\right)I_2^z - \left(\frac{d_2 t}{2}\right)I_{N-1}^z + \left(\frac{d_1-d_2}{2}\right)I_N^z$$
$$+ \frac{\omega_1}{8}S^x\left[(1-4I_2^z I_1^z) - 2(I_1^z - I_2^z)\right] + \frac{\omega_1}{8}I_1^x\left[(1-4I_3^z I_2^z) - 2(I_2^z - I_3^z)\right](1+2S^z)$$
$$+ \frac{1}{4}I_2^x(1-4I_3^z I_1^z)\left[(1-4I_4^z S^z) + 2(I_4^z - S^z)\right] + \frac{1}{4}I_{N-1}^x(1+2I_{N-3}^z)(1-4I_{N-2}^z I_N^z) \quad (19)$$
$$+ \frac{\omega_1}{8}I_N^x\left[(1-4I_{N-2}^z I_{N-1}^z) - 2(I_{N-1}^z - I_{N-2}^z)\right] + \frac{\omega_1}{4}\sum_{k=3}^{N-3}I_k^x(1-4I_{k-1}^z I_{k+1}^z)(1-4I_{k-2}^z I_{k+2}^z)$$

To estimate the efficiency of the one dimensional scheme we calculate the time dependence of the individual nuclear spin polarizations given by the definition

$$P_k(t) = Tr\{I_k^z e^{-itH}\rho(0)e^{itH}\} \quad (20)$$

The results of numerical calculation for the chain are identical to those obtained in [8] (see Figs. 1 and 2 of [8]). Note that the behavior of the second spin is critical. Irrespective of a chain length, flipping of the second spin will lead to propagation of a wave. Therefore, an analysis of the short spin chain with few numbers of spins give a trustworthy information of the wave propagation in the spin chains with large number of spins.

**b. Two-dimensional model**

The clusters of proton spins in monoclinic hydroxyapatite with the interchain dipolar couplings [15], can be used as 2D spin configuration for the measurement of the single spin state. The Hamiltonian of the $2D$ spin system, configuration of which is presented on Fig. 1b, has the following form

$$H = \frac{\omega_1}{2}\left(\sum_{m=1}^{N} I_m^x + S^x\right) + \sum_{q=1}^{M} d_q \sum_{m=1}^{N-1-q}\left[I_m^z I_{m+q}^z - \frac{1}{2}\left(I_m^x I_{m+q}^x + I_m^y I_{m+q}^y\right)\right] \qquad (21)$$
$$+ \sum_{m=1}^{N-1} f_m\left[I_m^z I_N^z - \frac{1}{2}\left(I_m^x I_N^x + I_m^y I_N^y\right)\right] + \sum_{m=1}^{N-1} g_m S^z I_m^z + g_N S^z I_N^z,$$

where $f_m$ is the dipolar coupling constants between of the $N$-th and $m$-th $I$ nuclear spins. In the approximation of nearest neighbors the effective Hamiltonian corresponding to the two dimensional scheme can be obtained

$$H_{eff} = \sum_{k=1}^{N-1}\left\{\left(\frac{f_k - g_k}{2}\right)I_k^z + \frac{\omega_1}{4}I_k^x\left[(1 - 4I_N^z S^z) - 2(I_N^z - S^z)\right]\right\}. \qquad (22)$$

The simulation results considering 7 spins $I = 1/2$ and a single spin, $S = 1/2$, in the weak coupling approximation are presented in Fig. 2a. Fig. 2b shows the dynamics of converting the polarization of the spin $S$ to the polarization of the ensemble of 7 spins $I = 1/2$ described by the Hamiltonian (21) at $\omega_1 = 0.15 D_1$ with the full dipole-dipole interactions.

**c. Three-dimensional Model**

For example, the sample contained labeled $^{13}$C-benzene is a good specimen of physical system of 3D configuration. The Hamiltonian of the $3D$ spin system, configuration of which is presented on Fig. 1c, includes the full dipole-dipole interactions between $I$ spins, has the following form

$$H = \frac{\omega_1}{2}\left(\sum_{m=1}^{N} I_m^x + S^x\right) + D_1 \sum_{q=1}^{M}\left[\frac{\sin\left(\frac{\pi}{N}\right)}{\sin\left(\frac{\pi q}{N}\right)}\right]^3 \sum_{m=1}^{N}\left[I_m^z I_{m+q}^z - \frac{1}{2}\left(I_m^x I_{m+q}^x + I_m^y I_{m+q}^y\right)\right]$$
$$+ \sum_{m=1}^{N-1} R_m\left[I_m^z I_N^z - \frac{1}{2}\left(I_m^x I_N^x + I_m^y I_N^y\right)\right] + \sum_{m=1}^{N-1} Q_m S^z I_m^z + Q_N S^z I_N^z. \quad (23)$$

where $R_m$ is the dipolar coupling constants between of the $N$-th and $m$-th $I$ nuclear spins, $Q_m$ is the dipolar coupling constants between of the $S$ and $m$-th $I$ nuclear spins, and where $Q_N$ is the dipolar coupling constants between of the $S$ and $N$-th $I$ nuclear spins In the approximation of nearest neighbors the effective Hamiltonian corresponding to the three dimensional scheme can be obtained

$$H = \frac{\omega_1}{2}\sum_{m=1}^{N-1} I_m^x \prod_{q=1}^{M}\left(1 - I_m^z I_{m+q}^z\right) + \sum_{m=1}^{N-1} R_m I_m^z I_N^z + \sum_{m=1}^{N-1} Q_m S^z I_m^z + Q_N S^z I_N^z. \quad (24)$$

Fig. 3a shows the dynamics of converting the polarization of the spin $S$ to the polarizations of the ensemble of 7 spins $I = 1/2$ in the weak coupling limit. Fig. 3b presents the converting process with the full dipole-dipole interactions, describing by the Hamiltonian (23) at $\omega_1 = 0.15 D_1$. At $\omega_1 = 0$, the spin diffusion wave [8] propagating from the first flipped spin, S, is not excited in the considered here 2D and 3D models because all spins have the same possibility to be flipped simultaneously. This wave does not provide any amplification since the total Z component of magnetization is conserved.

**4. Results and Discussion**

The spin dynamics is very different for the two different initial states: when the spin $S$ is in the initial state $|1\rangle$ - *up* or $|0\rangle$ - *down*. The state $|1\rangle$ of the spin $S$ does not change the states of the $I$ –spin system since it is an eigenstate of the effective Hamiltonian. While the state $|0\rangle$ - *down* can be converted into the states of the spins $I$. In the strong external field $H_0$, the secular part of the dipole-dipole interaction (4), includes interactions between all three spin components. In the one-dimensional linear spin chain, XX- and YY-interactions cause mutual flips of the spins even without the transverse resonant field ($\omega_1 = 0$) [8]. In this case, the first flipped spin propagates the spin-diffusion wave [8]. However, this wave does not give any amplification since the total Z-component of magnetization is conserved. The spin-diffusion processes in the 2D and 3D spin configurations are not excited since all spins have the same possibility to be flipped simultaneously and they are at the same initial state.

The most important feature of the spin dynamics described above is the signal amplification, when a state of polarization of a single spin $(S)$ is converted into a total polarization of the spin system $(I)$. The results for different models studied in this work are summarized in Fig.4. Fig. 4 shows the time dependences of the change in the total polarization, $\Delta P = P(t) - P(0)$ in $1D$, $2D$, and $3D$ spin system configurations. Here $P(t) = \sum_{k=1}^{N} P_k(t)$ is the total spin polarization. The potentiality of these schemes can be estimate according to the parameters: 1) coefficient of amplification, $\alpha = \frac{|\Delta P|}{2}$ [1]; 2) time of exposure, $T$ (in units of $1/\omega_1$), which necessary for achievement of the maximal contrast; and 3) exposure effectiveness, $\eta = \frac{C_{max}}{T}$, where $C_{max}$ is the maximal contrast. The coefficient of amplification, $\alpha$, and the exposure, $T$, have been extracted from Fig. 4. Results are presented in Table 1.

|  |  | Coefficient of amplification, $\alpha$ | Exposure, $T$ | Exposure effectiveness, $\eta$ | Contrast, $C$ |
|---|---|---|---|---|---|
| *1D* | Full d-d interaction | 1.85 | 8.91 | 0.21 | 1.23 |
|  | Weak coupling limit | 1.86 | 3.38 | 0.55 | 1.24 |
| *2D* | Full d-d interaction | 2.69 | 3.12 | 0.86 | 1.79 |
|  | Weak coupling limit | 2.88 | 3.15 | 0.91 | 1.92 |
| *3D* | Full d-d interaction | 0.89 | 3.63 | 0.24 | 0.59 |
|  | Weak coupling limit | 1.64 | 3.77 | 0.43 | 1.09 |

It is convenient to compare performance of different models by using the contrast [3,4] and the coefficient of amplification [1], defined as the ratio of the maximum total polarization change in the dynamics, triggered by a single spin flip, to the direct change of polarization resulted from a single spin flip. From the table we can see that the parameter which measures the efficiency of the amplification of the single spin state change with the dimensionality of the system. The shorter times of exposure, $T$, are achieved in 2D and 3D configuration of the spin system. This fact can be explained by taking into account that all spins in 2D and 3D models have the same possibility to be flipped simultaneously. The 1D and 2D configurations show larger coefficient of amplification, $\alpha$, and the contrast, $C$, than the 3D model used here. The system of linear chains used in 1D and 2D systems, only differs from the ring systems, used in 3D system, in the sense that there is no interaction of nuclear spins 1 and $N$ at the ends of the chains. This may be an indication that the amplification dynamics can be realized more effectively by using the linear chain as a part of amplification devices without interaction between spins of the ends. Besides that, the exposure effectiveness of amplification achieves the largest value for 2D model in which the both advantages are used, the potentialities of simultaneous spin flipping and the free ends of the spin chain. It is interesting that simulations with the weak coupling limit approximation give relative larger coefficient of amplification, $\alpha$, and the contrast, $C$, than simulation with the Hamiltonian (4), includes XX- and YY-interactions. For small spin clusters, long-range XX- and YY-interactions damage the amplified spin dynamics.

## 5. Conclusions

We studied the process of polarization conversion of a single nuclear spin to the total polarization of the ensemble of the nuclear spins in the *1D*, *2D*, and *3D* nuclear spin systems. It was demonstrated that efficient amplification dynamics can be organized for natural dipole-dipole couplings, both in the weak coupling limit and using the full dipole-dipole interactions. Numerical calculations were used to compare the potentiality of various spin models for the measurement of the single spin state.

**6. Acknowledgments**

The author thanks to J.-S. Lee and A. K. Khitrin for useful and stimulate discussions. This work was supported by the US-Israel Binational Science Foundation.

**Captions for figures**

Fig. 1. Configurations of a measuring device (the ensemble of 7 spins $I = 1/2$ (thin arrows) and the target spin S (thick arrow): a) $1D$ system, b) $2D$ system, and c) $3D$ system.

Fig. 2. (a) Time dependences of the conversion of the $S$ spin polarization to the individual spin polarizations in $2D$ system of seven dipolar-coupled spins calculated by using the Hamiltonian (21) ; (b) calculated by using the weak coupling limit.

Fig. 3. (a) Conversion of the $S$ spin polarization to the individual spin polarizations in $3D$ system of seven dipolar-coupled spins calculated by using the Hamiltonian (23); (b) calculated by using the weak coupling limit. Six of the curves are overlapping.

Fig. 4 The time dependences of the change in the total spin polarization, $\Delta P$ in $3D$ spin system configurations with the Hamiltonian (23) (solid) and for ZZ coupling (dash); and in $2D$ spin system with Hamiltonian (21) (dot) and for ZZ coupling (dash-dot); and in $1D$ spin system configurations for the alternating spin chain with dipole-dipole couplings (dash-dot-dot) and for ZZ coupling (shot-dash).

a)

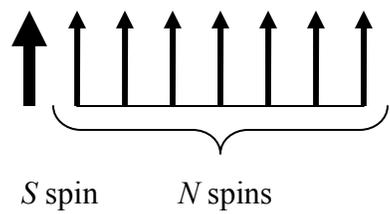

*S* spin    *N* spins

b)

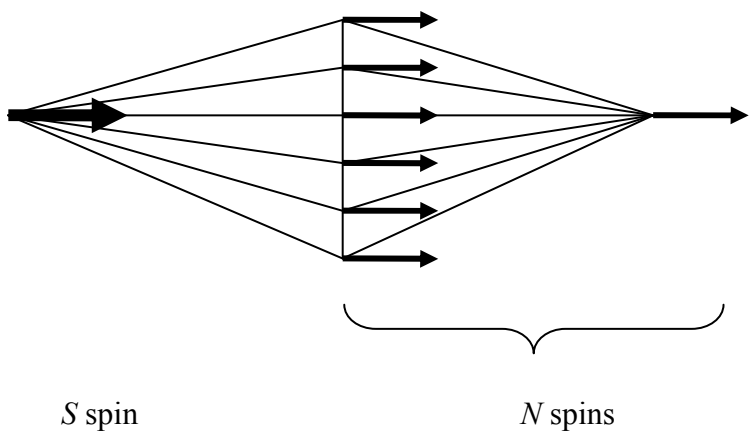

*S* spin    *N* spins

c)

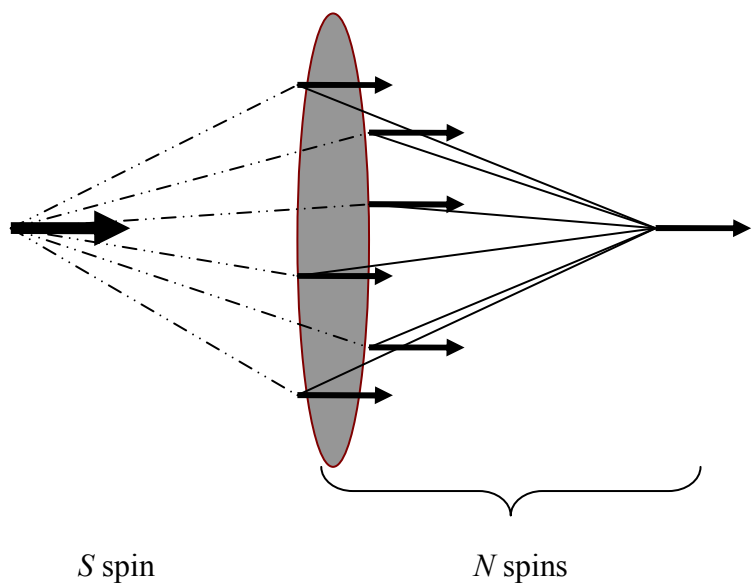

*S* spin    *N* spins

Fig. 1

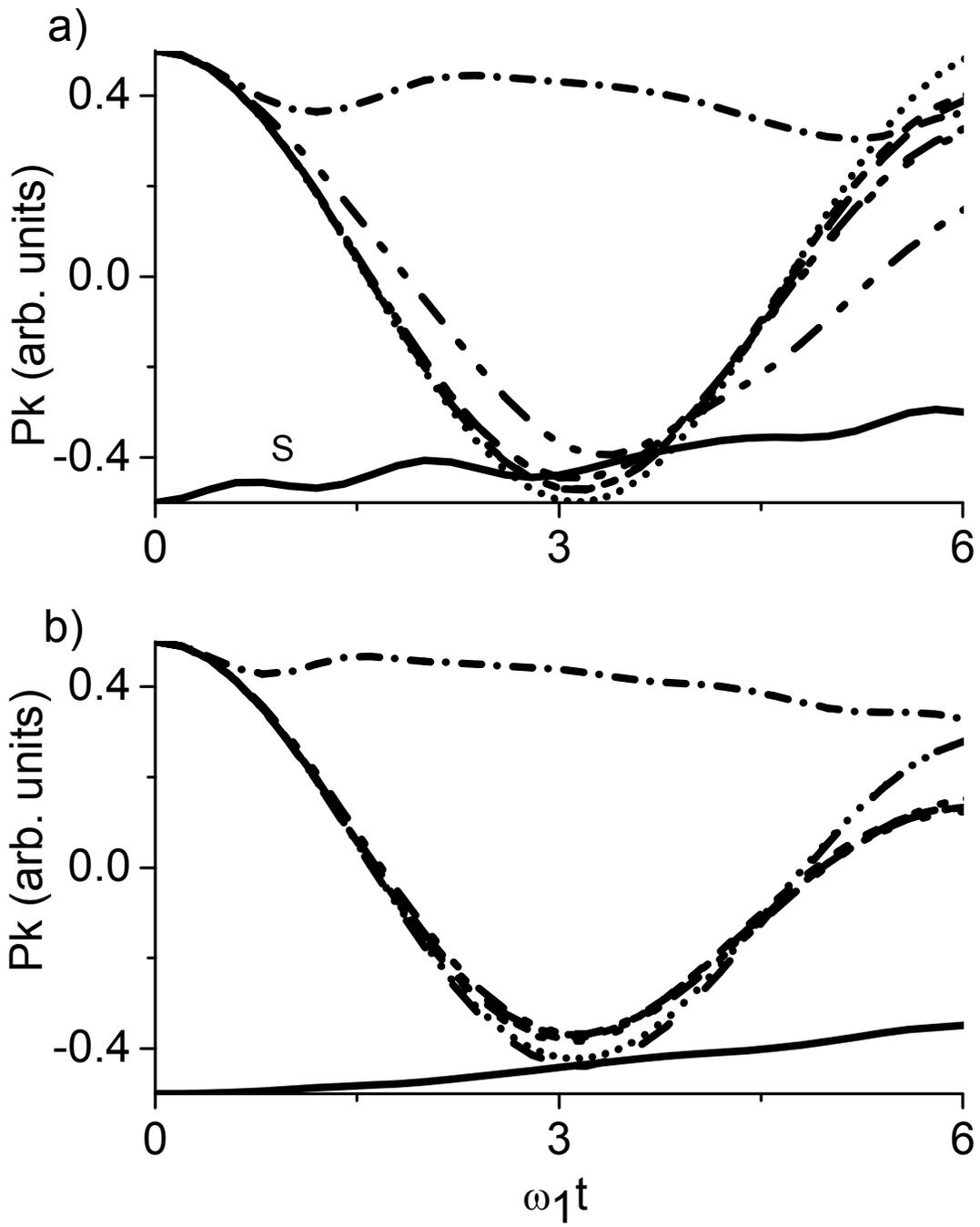

Fig. 2

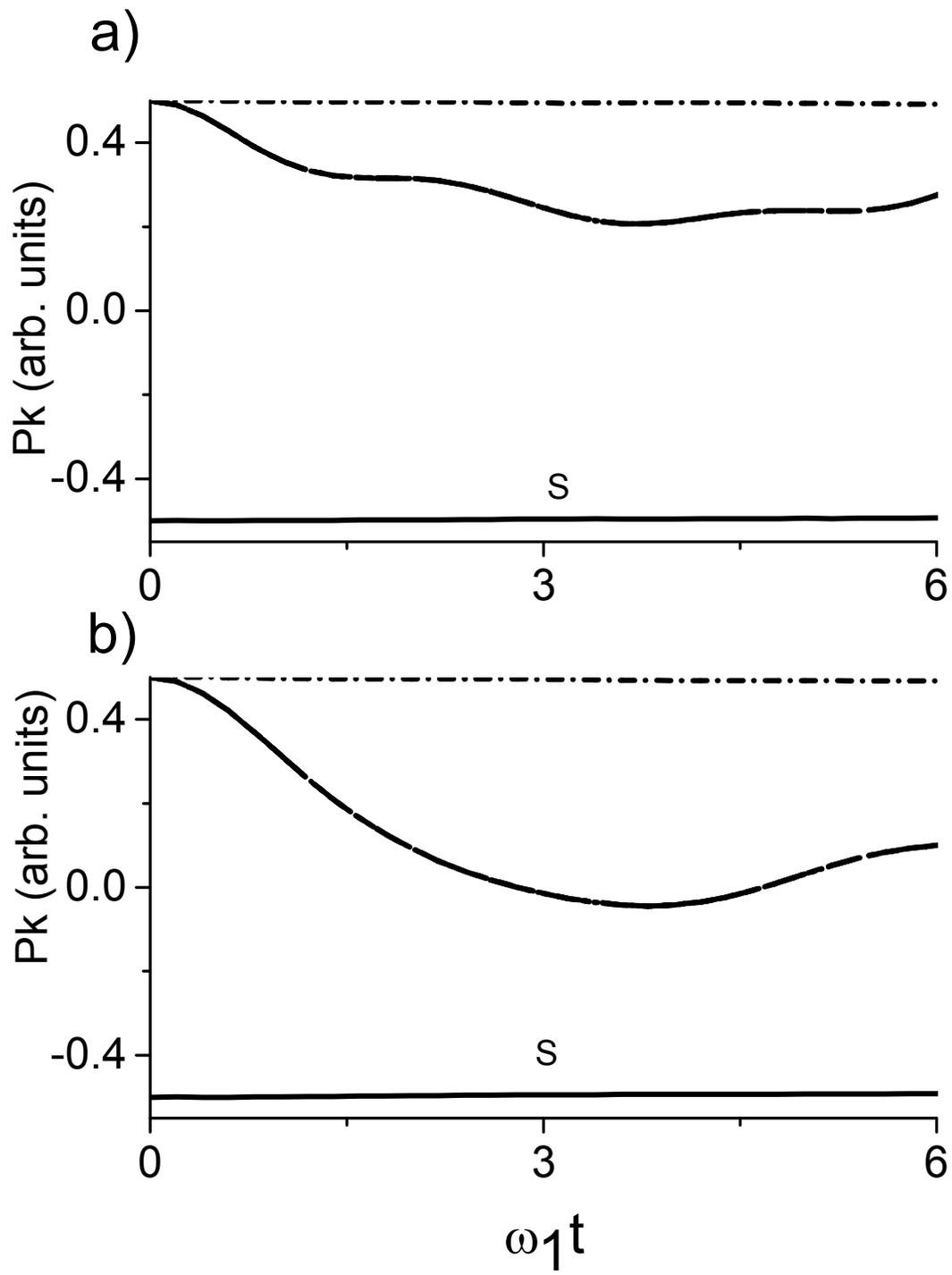

Fig.3

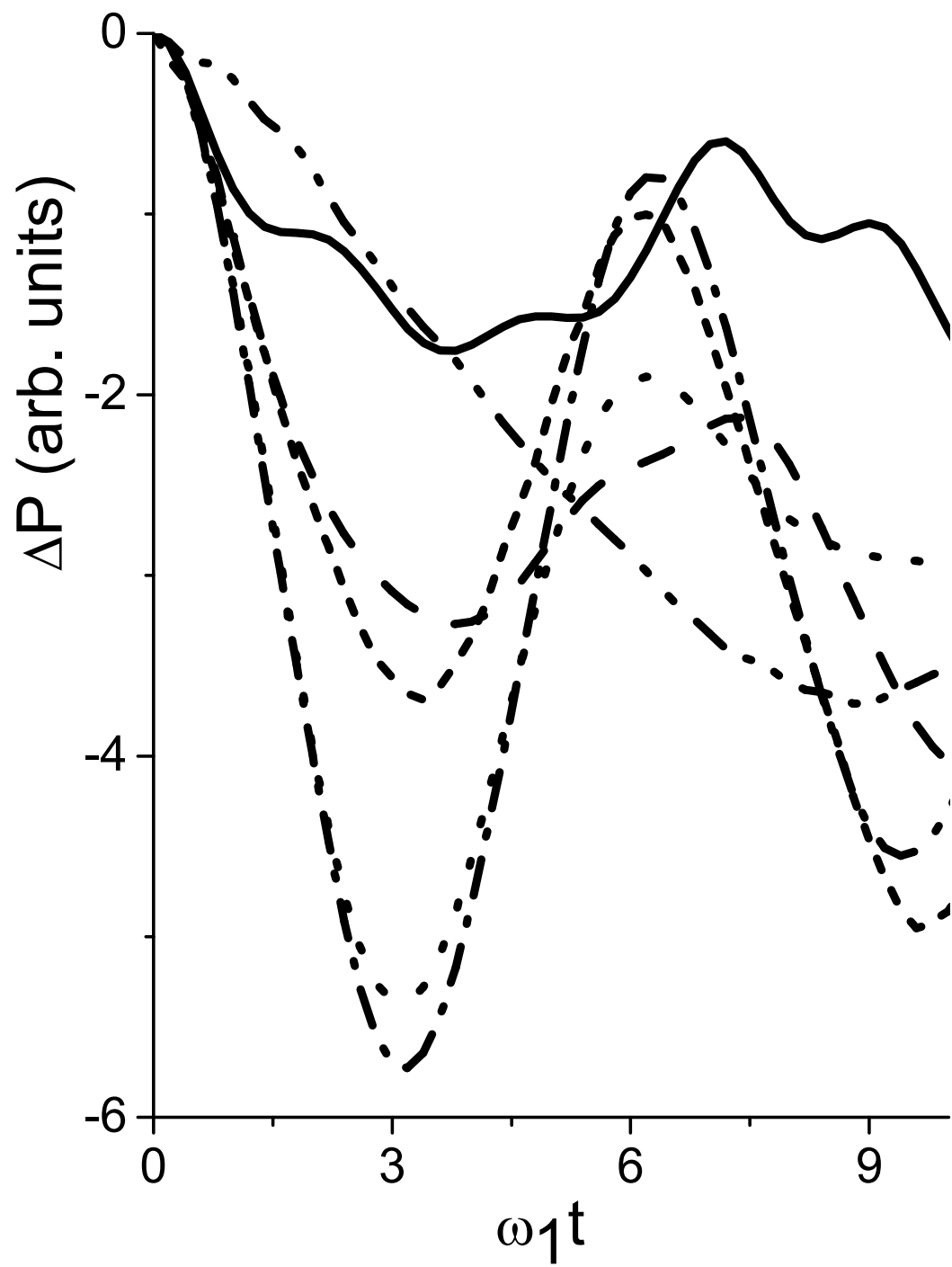

Fig.4